\documentclass[twocolumn,showpacs,superscriptaddress,preprintnumbers,amsmath,amssymb,aps,pre,10pt]{revtex4-1}

\usepackage{graphicx}
\usepackage{dcolumn}
\usepackage{bm}
\usepackage{color}

\begin{document}


\title{Generalized Entropy Theory of Glass Formation in Polymer Melts with Specific Interactions}

\author{Wen-Sheng Xu}
\email{wsxu@uchicago.edu}
\affiliation{James Franck Institute, The University of Chicago, Chicago, Illinois 60637, USA}

\author{Karl F. Freed}
\email{freed@uchicago.edu}
\affiliation{James Franck Institute, The University of Chicago, Chicago, Illinois 60637, USA}
\affiliation{Department of Chemistry, The University of Chicago, Chicago, Illinois 60637, USA}

\date{\today}

\begin{abstract}
Chemical structure has been long recognized to greatly influence polymer glass formation, but a general molecular theory that predicts how chemical structure determines the properties of glass-forming polymers has been slow to develop. While the generalized entropy theory (GET) explains the influence of various molecular details on polymer glass formation, the application of the GET has heretofore been limited to the use of the simplest polymer model in which all united atom groups within the monomers of a species interact with a common monomer averaged van der Waals energy. However, energetic heterogeneities are ubiquitous within the monomers of real polymers, and their implications for polymer glass formation remain to be investigated theoretically. This paper uses an extension of the GET to explore the influence of energetic heterogeneities within monomers upon the nature of polymer glass formation. This extension of the GET is achieved by combining the Adam-Gibbs theory relating the structural relaxation time to the configurational entropy with a recent significant extension of the lattice cluster theory for polymer melts with specific interactions, in particular, for melts where three distinct van der Waals interaction energies are required to describe the energetic heterogeneities within monomers. The present paper focuses on establishing general trends for the variation of characteristic properties of glass formation, such as the isobaric fragility parameter $m_P$ and the glass transition temperature $T_g$, with molecular details, such as the specific interactions and chain stiffness. Our computations confirm that the previously used model with monomer averaged interactions correctly captures general trends in the variation of $m_P$ and $T_g$ with various molecular parameters. More importantly, adjustment of the energetic heterogeneities within monomers alone are shown to provide an efficient mechanism for tailoring the properties of glass-forming polymers. The variations of polymer properties along iso-fragility and iso-$T_g$ lines are illustrated as important design tools for exhibiting the combined influence of specific interactions and chain stiffness.
\end{abstract}



\maketitle

\section{Introduction}

Despite the fact that polymeric materials readily form glasses upon cooling, a general theory for the nature of polymer glass formation and for the properties of polymer glasses remains elusive. Numerous studies demonstrate that molecular characteristics, such as chain stiffness, monomer structure, and the chemical structures of the backbone and side groups, profoundly affect polymer glass formation~\cite{Mac_26_6824, Mac_45_8430, Mac_31_4581, JCP_122_134505, SM_6_3430, Mac_41_7232,  PCCP_15_4604, PRL_97_045502, Mac_44_5044, Lan_29_12730, JCP_140_044901, JCP_142_074902}. Hence, understanding of the relation between material properties and these molecular details offers the potential to control polymer glass formation in a systematic manner. In particular, recent experiments~\cite{Mac_45_8430} indicate that both the glass transition temperature $T_g$ and the fragility parameter $m$, where $m$ measures the sensitivity of the structural relaxation time or viscosity to temperature changes, can be greatly tuned by modifying the chemical structure of the backbone and side groups and/or by controlling the spatial positions of the side groups with respect to the backbone and/or each other. Prior studies illustrate the important changes in properties that accompany variations in monomer structure and thus emphasize the importance of using knowledge concerning the dependence of polymer properties on molecular details to assist in the rational design of polymer materials. However, a complete fundamental understanding of how molecular details influence the properties of glass-forming polymers remains a challenge.

The generalized entropy theory (GET)~\cite{ACP_137_125, ACR_44_194} merges the lattice cluster theory (LCT)~\cite{ACP_103_335} for the thermodynamics of semiflexible polymers with the Adam-Gibbs (AG) theory~\cite{JCP_43_139, JCP_141_141102}, which invokes a relation between the structural relaxation time and the configurational entropy. Since the LCT employs an extended lattice model with structured monomers and describes the influence on polymer properties of short-range correlations imparted by chain connectivity, semiflexibility and interactions, the GET provides a convenient vehicle for systematically studying the changes introduced in polymer glass formation by the alterations of various molecular factors. While the agreement of the GET predictions with experiment for several nontrivial systems provides strong validation of the theory, the goal of rational design of polymeric materials requires considering the additional complexities of real polymer materials. In particular, all previous calculations within the GET~\cite{ACP_137_125, ACR_44_194, JPCB_109_21285, JPCB_109_21350, JCP_123_111102, JCP_124_064901, JCP_125_144907, JCP_131_114905, JCP_138_234501, Mac_47_6990} assume all united atom groups within an individual polymer species to interact with the same monomer averaged interaction energy. Although this monomer averaged interaction model suffices in establishing general trends observed in real polymers, real specific polymers generally contain energetic heterogeneities within monomers, i.e., different groups have disparate and specific interaction strengths. The implication of the influence on polymer properties and glass formation of energetic heterogeneities within monomers remains to be investigated within the GET.

Motivated by some recent experimental results of Sokolov and co-workers~\cite{Mac_45_8430} and by a desire to minimize the enormous complexity of the requisite LCT computations, we recently extended the LCT to treat a model of semiflexible polymers with the structures of poly($n$-$\alpha$-olefin) and with interactions where only the united atom groups at the end of side chains are assigned different nearest neighbor van der Waals interaction energies~\cite{JCP_141_044909}. Thus, three interaction energy parameters are used to describe the specific interactions in this new model. The focus of the present paper is to explore how the variation of the specific interactions can be used to exert greater control over the properties of glass-forming polymers. To this end, the new extension of the LCT is combined with the AG theory~\cite{JCP_43_139, JCP_141_141102}, thereby providing a similar generalization of the GET to describe polymer glass formation. Moreover, the greater realism introduced into the LCT and the GET by the new physical model enables testing the limits of validity of the simpler LCT model with a single monomer averaged van der Waals energy~\cite{ACP_137_125, ACR_44_194, JPCB_109_21285, JPCB_109_21350, JCP_123_111102, JCP_124_064901, JCP_125_144907, JCP_131_114905, JCP_138_234501, Mac_47_6990}, which is also discussed in detail in the present paper.

\section{Polymer Melts with Specific Interactions and Generalized Entropy Theory}

This section presents a brief introduction to necessary background information concerning the model of polymer melts with specific interactions and the GET of polymer glass formation. 

\subsection{Model Polymer Melts with Specific Interactions}

\begin{figure}[tb]
	\centering
	\includegraphics[angle=0,width=0.45\textwidth]{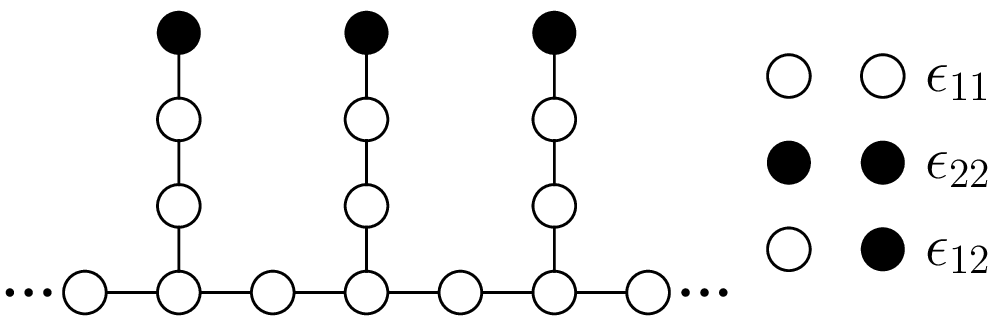}
	\caption{Lattice model considered here for the polymer chains in melts with specific interactions. Circles designate united atom groups, while lines represent the bonds between united atom groups. The chains are endowed with the structures of poly($n$-$\alpha$-olefin). The example depicted contains side groups each containing $n=3$ united atom groups. The united atom groups residing at the ends of each side chain have specific interactions, i.e., different nearest neighbor van der Waals interaction energies $\epsilon_{22}$ with each other and $\epsilon_{12}$ with all other united atom groups. Therefore, the model prescribes three different van der Waals interaction energy parameters ($\epsilon_{11}$, $\epsilon_{22}$ and $\epsilon_{12}$).}
\end{figure}

Our new lattice model considers polymer melts where the chains are endowed with the structures of poly($n$-$\alpha$-olefin) (Figure 1). The united atom groups residing at the ends of the side chains (represented as solid circles in Figure 1 and called e-groups in the following) differ from those lying on other positions (depicted by open circles in Figure 1 and termed n-groups in the following), since these side chains' end segments have different attractive nearest neighbor van der Waals interaction energies among themselves and with other n-groups. The microscopic van der Waals interaction energies $\epsilon_{11}$ and $\epsilon_{22}$ are assigned to the interaction of two nearest neighbor n-groups and two nearest neighbor e-groups, respectively, and the energy $\epsilon_{12}$ describes the interaction strength between a n-group and an e-group. This molecular structure is designed as a minimal model to account for the energetic heterogeneities within monomers of real polymers. These systems are thus termed polymer melts with specific interactions to distinguish them from polymer melts with monomer averaged interactions, where a single van der Waals energy specifies the interactions of the system. The model for polymer melts with specific interactions, of course, reduces to that with monomer averaged interactions when all nearest neighbor van der Waals interaction energies become identical, i.e., $\epsilon_{11}=\epsilon_{22}=\epsilon_{12}$.

We view this model of polymer chains with non-uniform intermolecular interactions as an idealization of polar polymers in which dipolar (or hydrogen bonding) species lie pendant off the chain backbone of the polymer chain and in which the remaining atomic species along the chain backbone interact with only van der Waals interactions. The range of dipole interactions should be limited in a medium having a relatively large dielectric constant, so that this model can be expected to capture some thermodynamic features of these polymer materials which have been of great interest with  regards to applications for energy storage devices~\cite{Mac_41_7232}. This model also serves as a minimal theoretical model for exploring the competition between the interactions between two atomic species on the nature of glass formation in the GET model, a matter of fundamental interest in understanding the nature of glass formation since real molecules ordinarily have such frustration in interactions.

In addition to the specific interactions, the LCT also includes a description of chain stiffness which is represented by the presence of a bending energy $E_b$. By construction, chains are fully flexible for $E_b=0$, while they become completely rigid in the limit $E_b\rightarrow\infty$. Also note that the side chains may have a different bending energy $E_s$ when the number of the united atom groups in each side chain $n$ exceeds or equals to two~\cite{JCP_124_064901}. Therefore, the free energy derived for the new model is a function of polymer volume fraction $\phi$, temperature $T$, interaction energies ($\epsilon_{11}$, $\epsilon_{22}$, $\epsilon_{12}$), bending energies ($E_b$, $E_s$), molecular weight $M$, and a set of geometrical indices that reflect the size, shape and bonding patterns of the monomers. The technical details of the theory and the explicit form for the LCT Helmholtz free energy are presented in ref~\cite{JCP_141_044909} for the model of polymer melts with specific interactions.

\subsection{Generalized Entropy Theory of Polymer Glass Formation}

The GET treats polymer glass formation as a broad transition with four characteristic temperatures~\cite{ACP_137_125}. These characteristic temperatures are evaluated from the LCT configurational entropy density $s_c$ (i.e., the configurational entropy per lattice site) at constant pressure. This $s_c$ exhibits a maximum as a function of $T$ at constant pressure, an essential feature for use in the AG model. Recent computations~\cite{JCP_141_234903} also indicate that the LCT configurational entropy density $s_c$ derived in ref~\cite{JCP_119_5730} is nearly identical to the ordinary entropy density $s=-\partial f/\partial T |_{\phi}$ (with $f$ designating the specific Helmholtz free energy) for the same thermodynamic conditions, probably because the lattice model is essentially devoid of vibrational contributions. Since the ordinary entropy density very closely approximates the configurational entropy density~\cite{JCP_141_234903} and since the former is much easier to calculate within the LCT, all calculations employ the ordinary entropy density in the present paper. For simplicity, the ordinary entropy density is just called the entropy density in the following.

The LCT computations for the temperature dependence of the entropy density $s(T)$ enable the direct determination of three characteristic temperatures of glass formation, namely, the ``ideal'' glass transition temperature $T_o$ where $s$ extrapolates to zero, the onset temperature $T_A$ which signals the onset of non-Arrhenius behavior of the relaxation time and which is found from the maximum in $s(T)$, and the crossover temperature $T_I$ which separates two temperature regimes with qualitatively different dependences of the relaxation time on temperature and which is evaluated from the inflection point in $Ts(T)$. The conventional definition of the fourth characteristic temperature, i.e., the glass transition temperature $T_g$, requires knowledge of the temperature dependence of the relaxation time $\tau$. For this purpose, the GET invokes the AG relation~\cite{JCP_43_139, JCP_141_141102},
\begin{equation}
	\tau=\tau_\infty\exp[\beta\Delta\mu s^\ast/s(T)],
\end{equation}
where $\tau_\infty$ is the high temperature limit of the relaxation time, $\beta=1/(k_BT)$ with being $k_B$ Boltzmann's constant, $\Delta\mu$ is the limiting temperature independent activation energy at high temperatures, and $s^\ast$ is the high temperature limit of $s(T)$. $\tau_\infty$ is set to be $10^{-13}$ s in the GET, which is a typical value for polymers~\cite{PRE_67_031507}. Motivated by the experimental data on the crossover temperature of various glass formers~\cite{PRE_67_031507}, the GET estimates the high temperature activation energy from the empirical relation $\Delta\mu=6k_BT_I$~\cite{ACP_137_125}. Thus, the relaxation time is computed within the GET without adjustable parameters beyond those used in the LCT for the thermodynamics of semiflexible polymers. The GET then identifies $T_g$ using the common empirical definition $\tau(T_g)=100$ s. In line with the original AG theory~\cite{JCP_43_139}, the relaxation time $\tau$ calculated from the GET reflects the slowest segmental relaxation process (i.e., the primary $\alpha$-relaxation process) in polymer fluids. Other relaxation processes, such as the local $\beta$-relaxation~\cite{Book_Ngai}, become the dominant relaxation below $T_g$, but they cannot yet be addressed in the GET.

Once the temperature dependence of the relaxation time is known, other related quantities are readily evaluated from the GET. For instance, the isobaric fragility parameter $m_P$ is determined from the standard definition~\cite{Science_267_1924},
\begin{equation}
	m_P=\left.\frac{\partial \log (\tau)}{\partial (T_g/T)}\right|_{P, T=T_g}.
\end{equation}

Illustrative computations of characteristic temperatures and fragility parameter appear in refs~\cite{ACP_137_125, ACR_44_194, Mac_47_6990}.

\section{Some Illustrative Computations}

This section first discusses the choices for the hetero-contact (or mixed) interaction energy $\epsilon_{12}$ for the model of polymer melts with specific interactions. The influence of the side group length on glass formation is then explored in order to demonstrate the physical validity of the LCT model for polymer melts with specific interactions. Finally, a correlation between the fragility parameter and the ratio of different characteristic temperatures, first revealed for the monomer averaged interaction model in ref~\cite{Mac_47_6990}, is shown to likewise apply for the specific interaction model.

\subsection{Influence of the Hetero-Contact Interaction Energy}

As discussed in Subsection II A, the model of polymer melts with specific interactions contains three interaction energy parameters ($\epsilon_{11}$, $\epsilon_{22}$ and $\epsilon_{12}$). Although the LCT allows the calculations to be performed with independent variations of each parameter, analyses of experimental data suggest that the hetero-contact interaction energy $\epsilon_{12}$ is subject to some constraints and cannot be completely independent of $\epsilon_{11}$ and $\epsilon_{22}$. For example, it is obviously impossible to find a real polymer where the hetero-contact interaction energy $\epsilon_{12}$ is significantly smaller or larger than either $\epsilon_{11}$ or $\epsilon_{22}$. Moreover, inspired by studies for polymer blends~\cite{JCP_140_194901, JCP_140_244905}, an exchange energy $\epsilon_\text{ex}$ can be similarly defined for the specific interaction model, 
\begin{equation}
	\epsilon_\text{ex}=\epsilon_{11}+\epsilon_{22}-2\epsilon_{12}.
\end{equation}
The physical essence of $\epsilon_\text{ex}$ for melts with specific interactions is of course different from that for blends. For instance, the sign of $\epsilon_\text{ex}$ is crucial for determining the nature of the phase behavior of a polymer blend~\cite{APS_183_63}. A positive exchange energy implies that the blend phase separates upon cooling. However, phase separation obviously cannot occur in a polymer melt. Nevertheless, it is instructive to explore whether the sign of $\epsilon_\text{ex}$ affects glass formation in the specific interaction model of polymer melts.

The above considerations suggest two different choices for considering the influence of the hetero-contact interaction energy $\epsilon_{12}$, i.e.,
\begin{equation}
	\epsilon_{12}=\sqrt{\epsilon_{11}\epsilon_{22}},
\end{equation}
and
\begin{equation}
	\epsilon_{12}=\epsilon_{11}+\epsilon_{22}-\sqrt{\epsilon_{11}\epsilon_{22}}.
\end{equation}
Both choices limit the number of free interaction energy parameters from three to two. The former rule (i.e., the Lorentz-Berthelot geometric mean approximation) guarantees that the exchange energy $\epsilon_\text{ex}$ is non-negative, while the latter sets $\epsilon_\text{ex}$ to be non-positive. We find that both choices produce very similar results of $T_g$ and $m_P$ (data not shown).

\begin{figure}[tb]
	\centering
	\includegraphics[angle=0,width=0.45\textwidth]{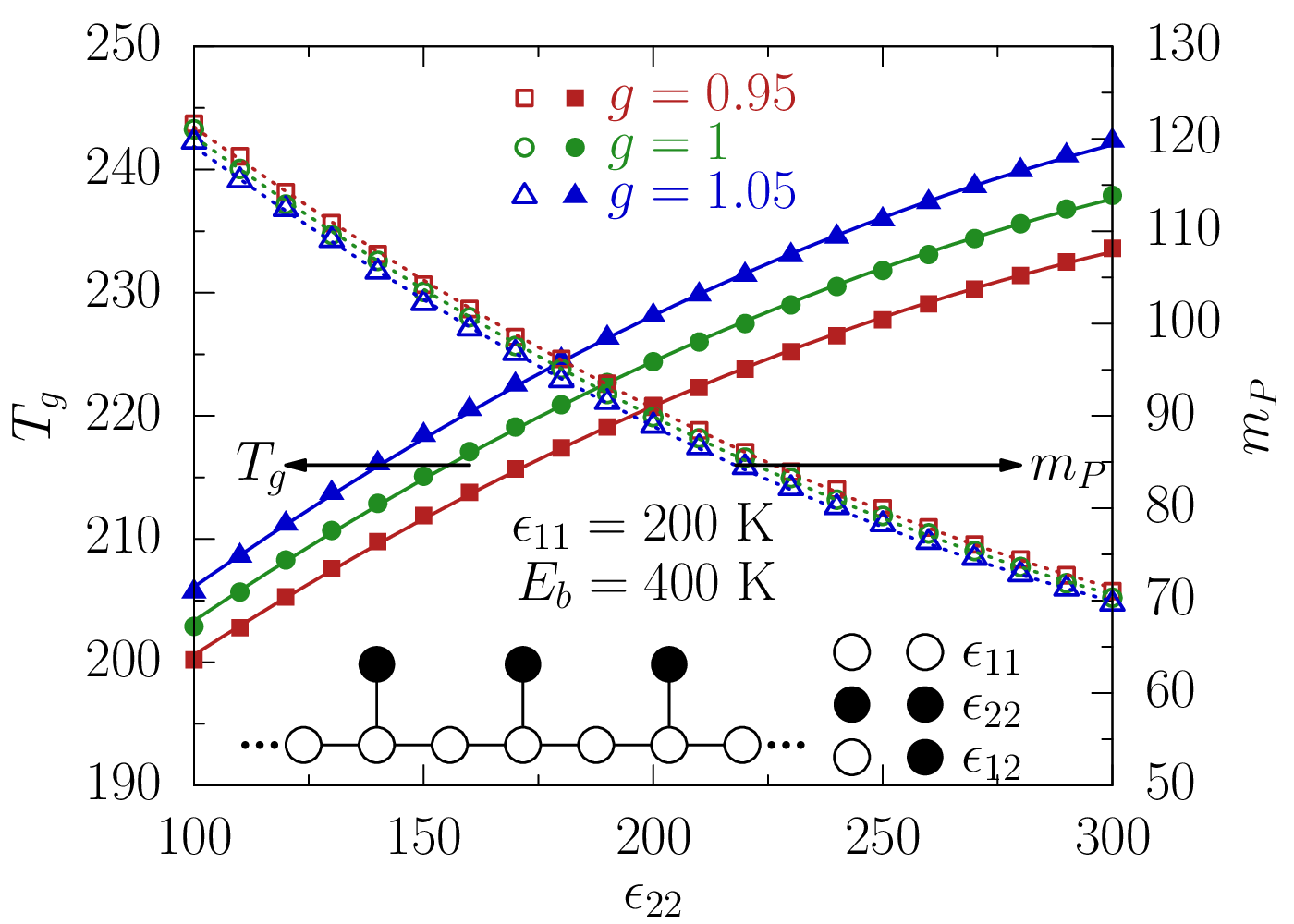}
	\caption{Glass transition temperature $T_g$ and isobaric fragility parameter $m_P$ as a function of $\epsilon_{22}$ for three values of $g$. The computations assume a constant pressure of $P=1$ atm, and the chains possess the structure of poly(propylene) (PP) with molecular weight $M=24001$, $\epsilon_{11}=200$ K, and bending energy $E_b=400$ K. The lines are a guide to the eye. The inset depicts the lattice model of polymer melts with specific interactions and with the structure of PP, a model which is extensively used in the present work.}
\end{figure}

The recent work of Lipson and co-workers~\cite{Mac_45_1076, Mac_45_8861} demonstrates that a modified geometric mean approximation for $\epsilon_{12}$,
\begin{equation}
	\epsilon_{12}=g\sqrt{\epsilon_{11}\epsilon_{22}},
\end{equation}
provides a better description for understanding the phase behavior of polymer blends, where the parameter $g$ characterizes the departure of $\epsilon_{12}$ with respect to the geometric mean. For instance, a polymer blend is suggested to exhibit a lower critical solution temperature for $g>1$, while an upper critical solution temperature is more likely to occur for $g<1$~\cite{Mac_45_1076, Mac_45_8861}. Therefore, it is interesting to explore the influence of $\epsilon_{12}$ given by eq 6 on glass formation in polymer melts with specific interactions.

We consider polymer melts of chains with the structure of poly(propylene) (PP, where the side group length is $n=1$) where a single bending energy for the backbone $E_b$ describes the flexibility of the chains. All calculations in the present paper use the same parameter set: the lattice coordination number is $z=6$; the cell volume parameter is $a_{\text{cell}}=2.7$\AA{}; the pressure is $P=1$ atm; and the molecular weight (i.e., the total number of united atom groups in a single chain) is chosen to be $M=24001$, corresponding to a polymer melt of chains with high molecular weight. Figure 2 displays $T_g$ and $m_P$ as a function $\epsilon_{22}$ for $g=0.95$, $1$ and $1.05$~\cite{note1}, when $\epsilon_{11}$ is fixed to be $200$ K. An increase in $g$ leads to an elevated $T_g$ and a slightly diminished $m_P$. Nevertheless, the qualitative trends of both $T_g$ and $m_P$ are not affected by $g$. Thus, changing the sign of $(g-1)$ has merely quantitative effects on both $T_g$ and $m_P$ for the specific interaction model. Therefore, the subsequent calculations employ the geometric mean approximation (i.e., eq 4) to fix the value of $\epsilon_{12}$, with the recognition that other choices would yield similar results. The influence of the interaction energy parameters on both $T_g$ and $m_P$ is analyzed in detail in Secs. IV and V.

\subsection{Influence of the Side Group Length}

\begin{figure}[tb]
	\centering
	\includegraphics[angle=0,width=0.45\textwidth]{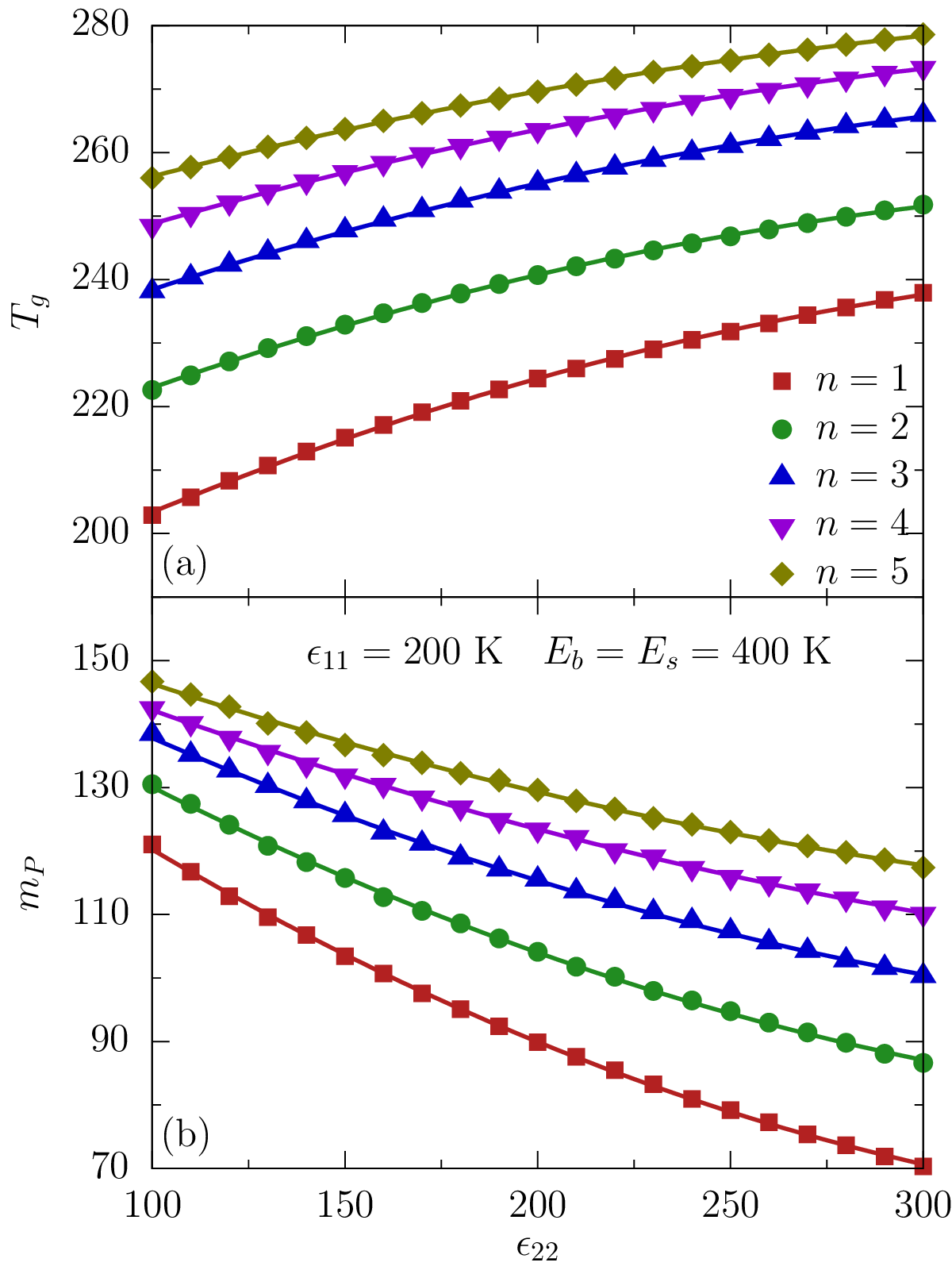}
	\caption{(a) Glass transition temperature $T_g$ and (b) isobaric fragility parameter $m_P$ as a function of $\epsilon_{22}$ for various side group lengths $n$. The computations are performed for polymer melts with specific interactions at a constant pressure of $P=1$ atm, where the chain molecular weight is $M=24001$, the other interaction energies are $\epsilon_{11}=200$ K and $\epsilon_{12}=\sqrt{\epsilon_{11}\epsilon_{22}}$, and the bending energies for backbone and side chains are $E_b=E_s=400$ K.}
\end{figure}

Previous studies demonstrate that the side group length significantly influences the glass-formation properties in polymer melts with monomer averaged interactions~\cite{ACP_137_125, JCP_131_114905, ACR_44_194, JCP_138_234501}, implying that controlling side group structure provides a powerful means to regulate the properties of glass-forming polymers. Figure 3 displays both $T_g$ and $m_P$ as strongly depending on the side group length $n$ in polymer melts with specific interactions. The influence of $n$ on $T_g$ and $m_P$ is even stronger than that of interaction energy parameters; e.g., the magnitude of $T_g$ at $\epsilon_{22}=100$ K for $n=5$ is larger than that at $\epsilon_{22}=300$ K for $n=1$. Another noticeable feature in Figure 3 is that the change of $T_g$ and $m_P$ with $\epsilon_{22}$ weakens for larger $n$. This result is not surprising since the fraction of the side chains' end segments in a single chain varies with $n$ roughly as $f_e=1/(n+2)$, indicating that the influence of $\epsilon_{22}$ is less important for larger $n$. Thus, Figure 3 also provides evidence for the physical validity of the LCT free energy derived for polymer melts with specific interactions.

Notice that $T_g$ increases as the side group length $n$ grows for fixed $\epsilon_{22}$ in Figure 3a, a trend that arises because the illustrative calculations in Figure 3 employ the flexible-flexible (F-F) polymer model~\cite{ACP_137_125}, where both chain backbone and side groups are flexible ($E_b=E_s=400$ K). A trend of increasing $T_g$ with $n$ occurs also for flexible-stiff (F-S) polymers with $E_b<E_s$~\cite{ACP_137_125}, a trend in accord with measurements for the pair  polystyrene (PS)~\cite{Mac_45_8430} and poly($2$-vinyl naphthalene) (P2VN)~\cite{JPSA_40_583}, systems with fairly rigid and extended side groups. Specifically, experiments indicate that $T_g$ for P2VN exceeds that for PS ($T_g=372$ K for PS)  by $50 $ K. The GET predicts that the other class of polymers, i.e., stiff-flexible (S-F) polymers with $E_b>E_s$, behave in the opposite fashion where elevating $n$ leads to a diminished $T_g$,~\cite{ACP_137_125, JCP_131_114905} a trend that has been found in polymers with stiff backbone and flexible side groups; e.g., $T_g$ is shown in ref~\cite{Mac_31_6951} to decrease significantly with increasing side group length for the homologous series of poly($n$-alkyl methacrylates), polymers with chemical structures that are typical of S-F polymers. Therefore, the GET predicted trends for the $n$-dependence of $T_g$ are generally in agreement with the observations in real polymers.

\subsection{Correlation between Fragility Parameter and Breadth of Glass Formation}

\begin{figure}[tb]
	\centering
	\includegraphics[angle=0,width=0.45\textwidth]{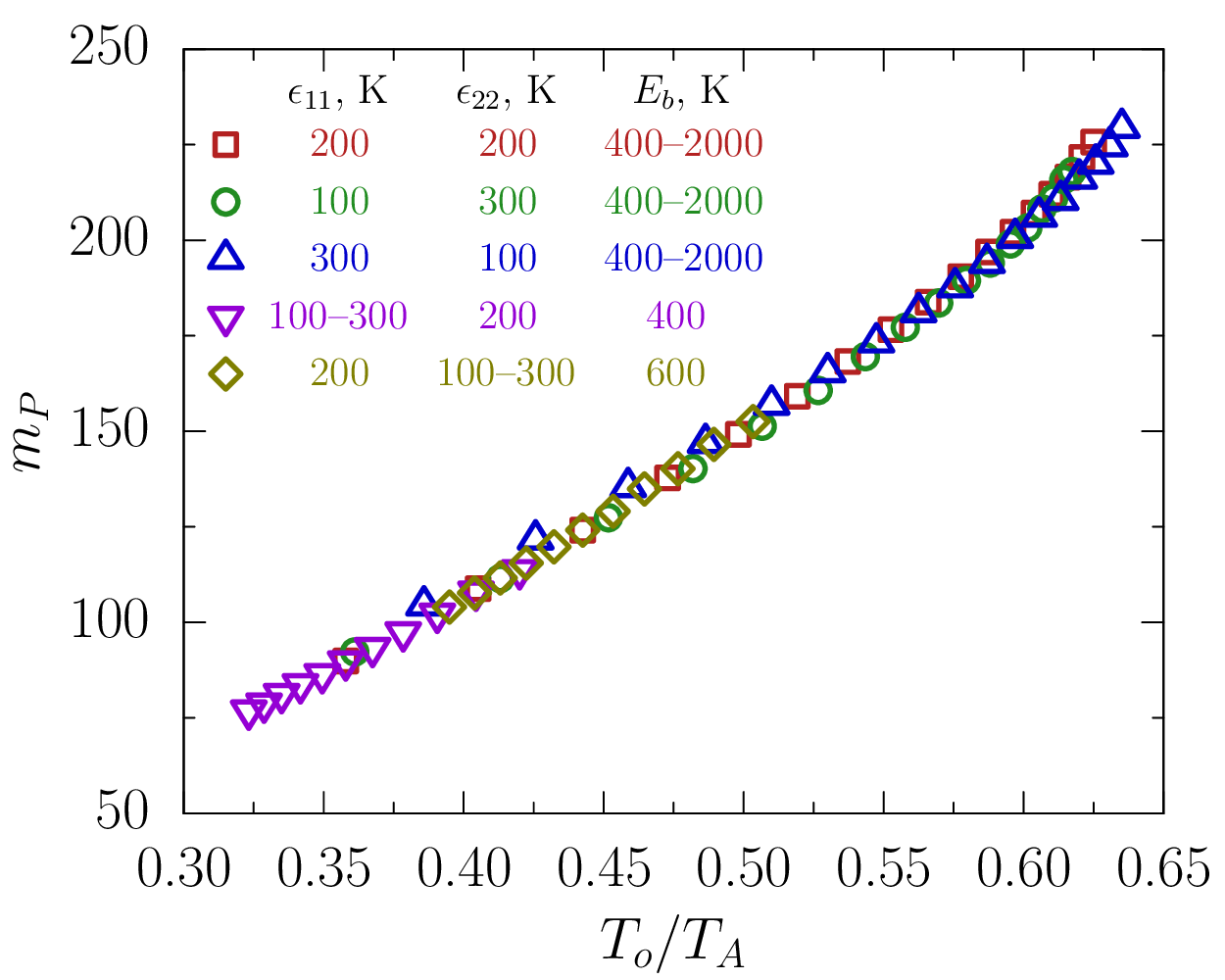}
	\caption{Correlation between the isobaric fragility parameter $m_P$ and the characteristic temperature ratio $T_o/T_A$. The computations consider polymer melts with specific interactions at a constant pressure of $P=1$ atm, where the chains possess the structure of PP with the molecular weight $M=24001$. The range for each parameter is indicated in the legend.}
\end{figure}

A previous work~\cite{Mac_47_6990}, invoking the monomer averaged interaction model, demonstrates that a master curve exists between the isobaric fragility parameter $m_P$ and the characteristic temperature ratios such as $T_o/T_A$. This result indicates that the commonly used $m_P$ indeed correlates with ratios of the characteristic temperatures, supporting the contention that the breadth of the glass-formation process provides a promising measure for the fragility of glass-forming liquids.~\cite{ACP_137_125} Here, we test whether the specific interaction model exhibits similar features. Figure 4 displays the correlation between $m_P$ and $T_o/T_A$ for independent variations of each parameter ($\epsilon_{11}$, $\epsilon_{22}$, or $E_b$). The collapse of all data, including those from the monomer averaged interaction model (i.e., squares in Figure 4), indicates that the same master curve applies equally for the monomer averaged interaction model~\cite{Mac_47_6990} and for the model of polymer melts with specific interactions. Therefore, Figure 4 suggests that the fragility generally correlates with the breadth of glass formation in glass-forming polymers.

Next, we focus on how molecular factors control the properties of glass formation of melts with specific interactions. Specifically, sections 4 and 5 demonstrate that both $m_P$ and $T_g$ can be finely tailored by altering the molecular parameters, such as interaction energies and chain stiffness. Since the simplest model for poly($n$-$\alpha$-olefins) with $n>1$ contain separate bending energies for the backbone and side groups, to simplify the discussion, the following calculations only consider chains with the structure of PP because this choice requires the minimal number of parameters in the LCT~\cite{JCP_124_064901}. Given the chain structure as that of PP, the microscopic interaction energies $\epsilon_{11}$, $\epsilon_{22}$, and $\epsilon_{12}$ precisely correspond to backbone-backbone, side group-side group, and backbone-side group interaction energies.

\section{Controlling Polymer Glass Formation by Altering Side Group-Side Group Interactions and Chain Stiffness}

This section illustrates how the properties of glass-forming polymer melts can be systematically controlled by adjusting side group-side group interactions and chain stiffness. Since a recent work~\cite{Mac_47_6990} investigates the combined influence of similar molecular parameters for the model of polymer melts with monomer averaged interactions, we test whether the qualitative features found for the monomer averaged interaction model apply for the more realistic model with specific interactions. This section also discusses the behavior of properties along the iso-fragility and iso-$T_g$ lines in the plane of the side group-side group interaction energy and bending energy, concepts first introduced in ref~\cite{Mac_47_6990}.

The backbone-backbone interaction energy remains at $\epsilon_{11}=200$ K for the next set of calculations and hence, the side group-side group interaction energy $\epsilon_{22}$ and bending energy $E_b$ are the control parameters. This class of models can be regarded as polymers with similar chemical backbone structures but different side groups and is motivated by the recent experimental results of Sokolov and co-workers~\cite{Mac_45_8430}, who explore the alteration of polymer glass formation with changes in polar interactions by comparing the properties for polymers with the same backbone but different side groups.  For example, PP, poly(vinyl alcohol) (PVA), and poly(vinyl chloride) (PVC) have the same backbone but different side groups. Comparing their behavior of glass formation leads to a better understanding of how chemical structure controls the properties of glass-forming polymers~\cite{Mac_45_8430}.

\subsection{Combined Influence of Side Group-Side Group Interactions and Chain Stiffness}

\begin{figure}[tb]
	\centering
	\includegraphics[angle=0,width=0.45\textwidth]{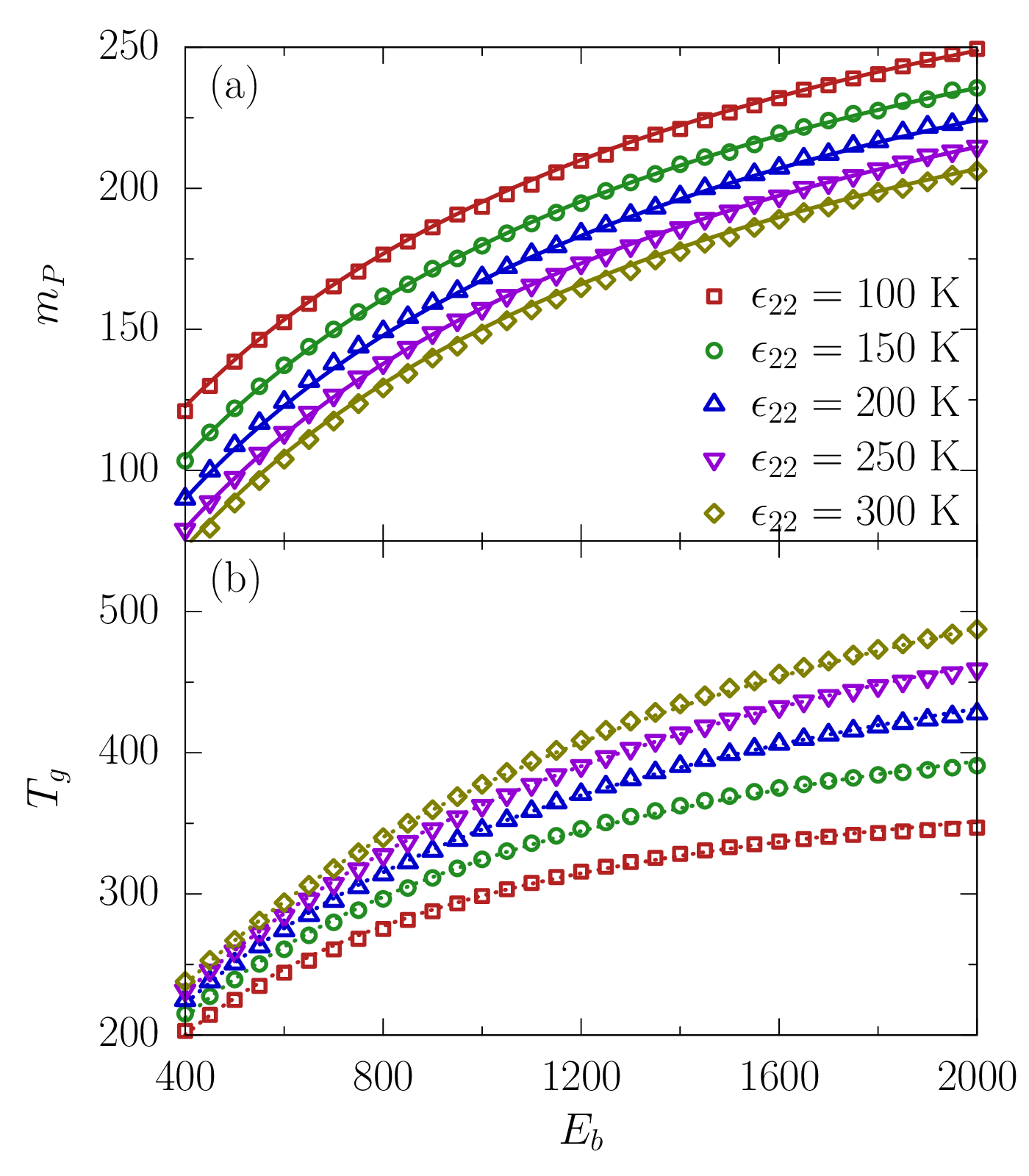}
	\caption{(a) Isobaric fragility parameter $m_P$ and (b) glass transition temperature $T_g$ as a function of the bending energy $E_b$ for various side group-side group interaction energies $\epsilon_{22}$. Solid lines (a) and dotted lines (b) are fits according to eqs 7 and 8 with the parameters $a_0=82.5421$, $a_1=-0.740738$, $a_2=1.10301\times10^{-3}$, $b_0=0.412431$, $b_1=-1.45264\times10^{-4}$, $c_0=1.11504\times10^{-3}$, $c_1=2.34999\times10^{-7}$ and $u_0=88.4981$, $u_1=-1.07334\times10^{4}$, $u_2=-3.25406\times10^{5}$, $v_0=0.556738$, $v_1=30.3834$, $v_2=3.30207\times10^{3}$, $w_0=3.28156\times10^{-4}$, $w_1=0.16638$, $w_2=8.24995$.}
\end{figure}

Figure 5 displays the isobaric fragility parameter $m_P$ and the glass transition temperature $T_g$ as a function of bending energy $E_b$ for various side group-side group interaction energies $\epsilon_{22}$. Both $m_P$ and $T_g$ increase with $E_b$ and tend to saturate for very large $E_b$ at fixed $\epsilon_{22}$. The trends of elevating $m_P$ and $T_g$ with $E_b$ are generally in accord with physical intuition and experimental observations indicating that greater chain rigidity leads to larger fragilities and higher $T_g$. For example, $m_P$ and $T_g$ for a stiff polymer, such as poly(t-butylstyrene) ($m_P=141$ and $T_g=407$ K),~\cite{Mac_41_7232} can greatly exceed those for a more flexible polymer, such as poly(dimethylsiloxane) ($m_P=85$ and $T_g=143$ K)~\cite{Mac_41_7232}. Increasing $\epsilon_{22}$, however, leads to a drop in $m_P$ but an elevation in $T_g$ at the same $E_b$. These qualitative trends are the same as observed for the monomer averaged interaction model~\cite{JCP_131_114905, Mac_47_6990}, providing evidence that the monomer averaged interaction model correctly captures general trends in the variation of both quantities with interaction energy and chain stiffness. Moreover, two simple algebraic equations fairly accurately capture the computed combined variations of $m_P$ and $T_g$ with $\epsilon_{22}$ and $E_b$,
\begin{equation}
	m_P=\frac{a_0+a_1\epsilon_{22}+a_2\epsilon_{22}^2+(b_0+b_1\epsilon_{22})E_b}{1+(c_0+c_1\epsilon_{22})E_b}, 
\end{equation}
\begin{equation}
	T_g=\frac{u_0+u_1/\epsilon_{22}+u_2/\epsilon_{22}^2+(v_0+v_1/\epsilon_{22}+v_2/\epsilon_{22}^2)E_b}{1+(w_0+w_1/\epsilon_{22}+w_2/\epsilon_{22}^2)E_b}, 
\end{equation}
where the fitted parameters $a_{\alpha}(\alpha=0,...,2)$, $b_{\alpha}(\alpha=0,1)$, $c_{\alpha}(\alpha=0,1)$, $u_{\alpha}(\alpha=0,...,2)$, $v_{\alpha}(\alpha=0,...,2)$ and $w_{\alpha}(\alpha=0,...,2)$ are summarized in the caption of Figure 5. The forms of the equations are essentially the same as those proposed for analyzing the influence of monomer averaged interaction $\epsilon$ and bending $E_b$ energies on $m_P$ and $T_g$ for the model of melts with monomer averaged interactions~\cite{Mac_47_6990}. The fitting functions are chosen to ensure the observed saturation of $m_P$ and $T_g$ for large $E_b$.

\begin{figure}[tb]
	\centering
	\includegraphics[angle=0,width=0.45\textwidth]{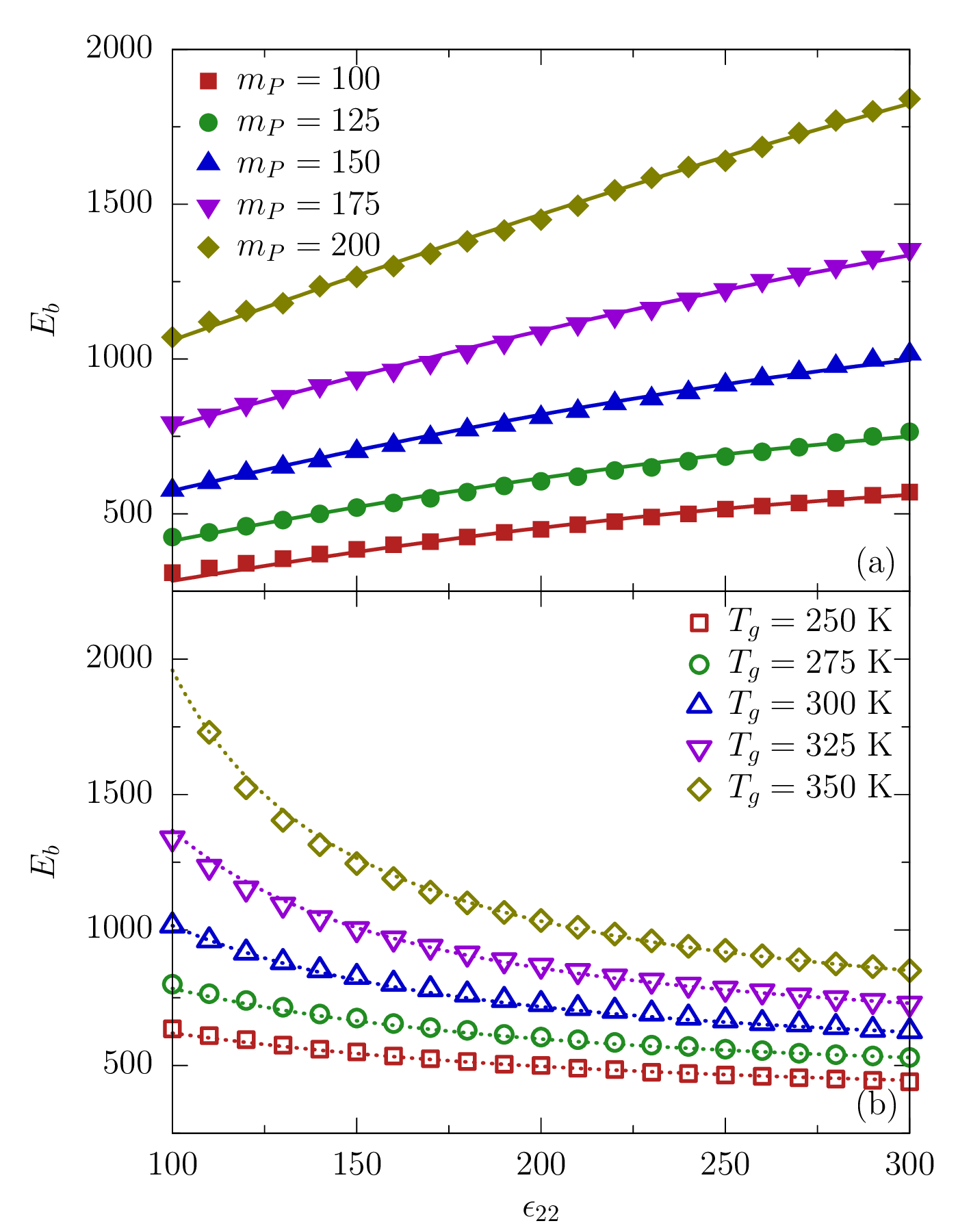}
	\caption{(a) Iso-fragility lines and (b) iso-$T_g$ lines in the plane of side group-side group interaction energy $\epsilon_{22}$ and bending energy $E_b$ for several representative values of $m_P$ and $T_g$. Solid lines (a) and dotted lines (b) are the results of eqs 7 and 8 with the fitted parameters given in the caption of Figure 5.}
\end{figure}

Following the previous work~\cite{Mac_47_6990}, we define two types of special lines in the $\epsilon_{22}$-$E_b$ plane, along which either the fragility parameter $m_P$ or the glass transition temperature $T_g$ remains constant. Their existence is due to the combined influence of $\epsilon_{22}$ and $E_b$ on $m_P$ and $T_g$. These lines are termed iso-fragility and iso-$T_g$ lines, respectively, and have been analyzed extensively for the monomer averaged interaction model in order to better understand some experimental results of Sokolov and co-workers~\cite{Mac_45_8430}.

Figure 6 displays several iso-fragility and iso-$T_g$ lines as curves of $\epsilon_{22}$ vs. $E_b$ for representative values of $m_P$ and $T_g$, respectively. The lines in Figure 6 display the fits from eqs 7 and 8 with the parameters given in the caption of Figure 5. In line with the trends observed for the monomer averaged interaction model~\cite{Mac_47_6990}, Figure 6a indicates that $E_b$ grows approximately linearly with $\epsilon_{22}$ along an iso-fragility line and that the slope grows with $m_P$, and Figure 6b exhibits the trend that $E_b$ decreases with $\epsilon_{22}$ along the iso-$T_g$ lines, so chains must be more flexible at larger cohesive energies in order for the system to achieve the same $T_g$. The trends in Figure 6 can be understood as dictated by the combined influence of $\epsilon_{22}$ and $E_b$ on $m_P$ and $T_g$. The properties along the iso-fragility and iso-$T_g$ lines are analyzed in subsection 4.2.

\subsection{Properties along Iso-Fragility and Iso-$T_g$ Lines}

Our previous analysis~\cite{Mac_47_6990} for the monomer averaged interaction model demonstrates that many properties, e.g., the entropy density, the polymer volume fraction and the relaxation time at characteristic temperatures, such as $T_g$, remain invariant along the iso-fragility lines in the $\epsilon$-$E_b$ plane. By contrast, no special characteristics are found along the iso-$T_g$ lines in the $\epsilon$-$E_b$ plane. Therefore, the concept of fragility appears to provide more fundamental insight into glass formation than the glass transition temperature $T_g$. Here, we test whether the model of polymer melts with specific interactions displays the same properties.

\begin{figure}[tb]
	\centering
	\includegraphics[angle=0,width=0.45\textwidth]{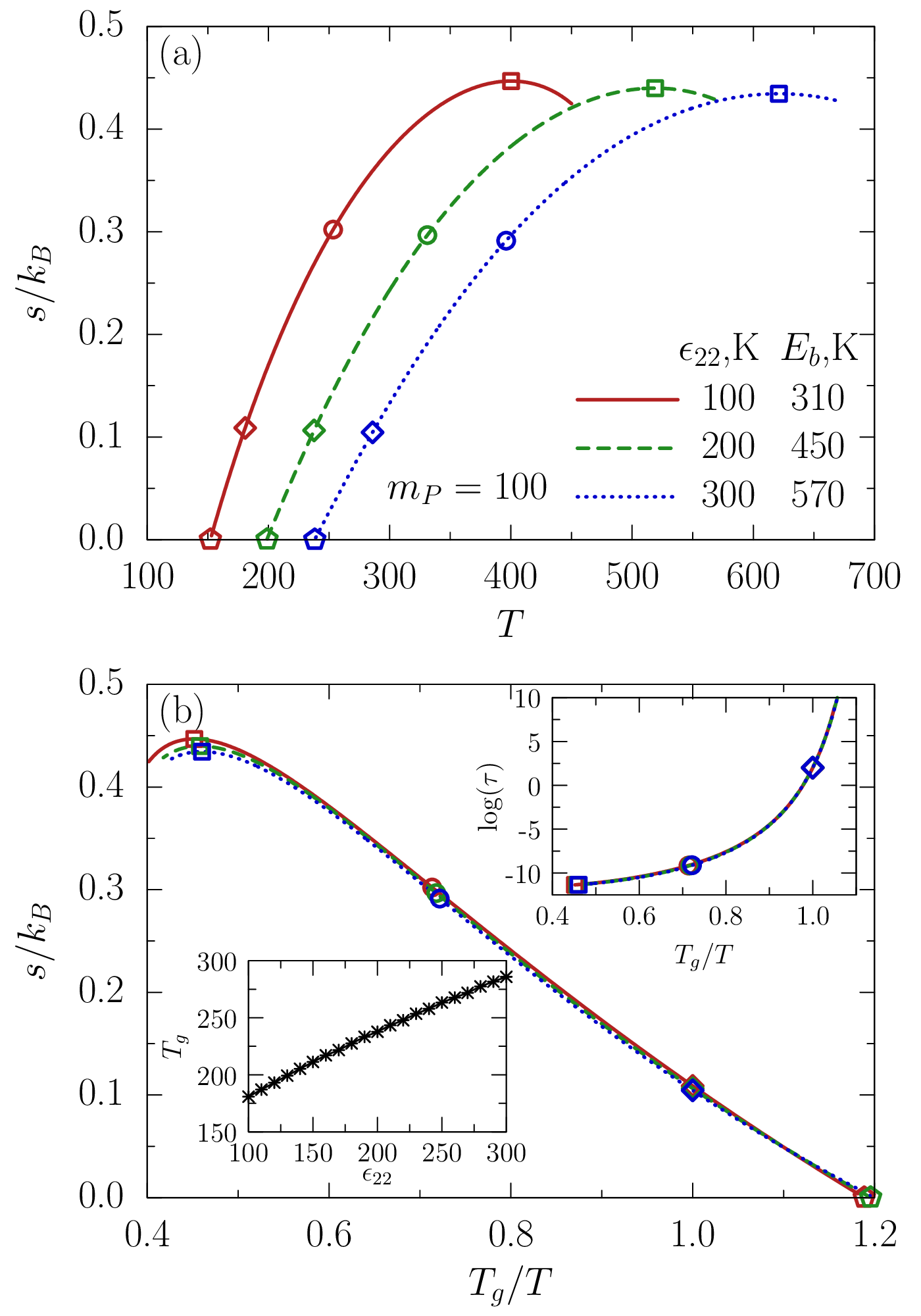}
	\caption{(a) Entropy density $s/k_B$ as a function of $T$ for several pairs of $\epsilon_{22}$ and $E_b$ that produce the same isobaric fragility parameter of $m_P=100$.  (b) $T_g$-scaled Arrhenius plot for the entropy density $s/k_B$ for the same pairs of $\epsilon_{22}$ and $E_b$ as in part a. The upper inset to part b presents the $T_g$-scaled Arrhenius plot for the relaxation time $\tau$ for the same pairs of $\epsilon_{22}$ and $E_b$ as in  part a, while the lower inset to part b depicts $\epsilon_{22}$-dependence of $T_g$ along an iso-fragility line for $m_P=100$. Squares, circles, diamonds and pentagons designate the positions of characteristic temperatures $T_A$, $T_I$, $T_g$ and $T_o$, respectively. Note that the vanishing of the entropy density is probably an artifact of high $T$ expansion in the LCT; see refs~\cite{JCP_125_144907, JCP_141_234903} for discussion.}
\end{figure}

\begin{figure}[tb]
	\centering
	\includegraphics[angle=0,width=0.45\textwidth]{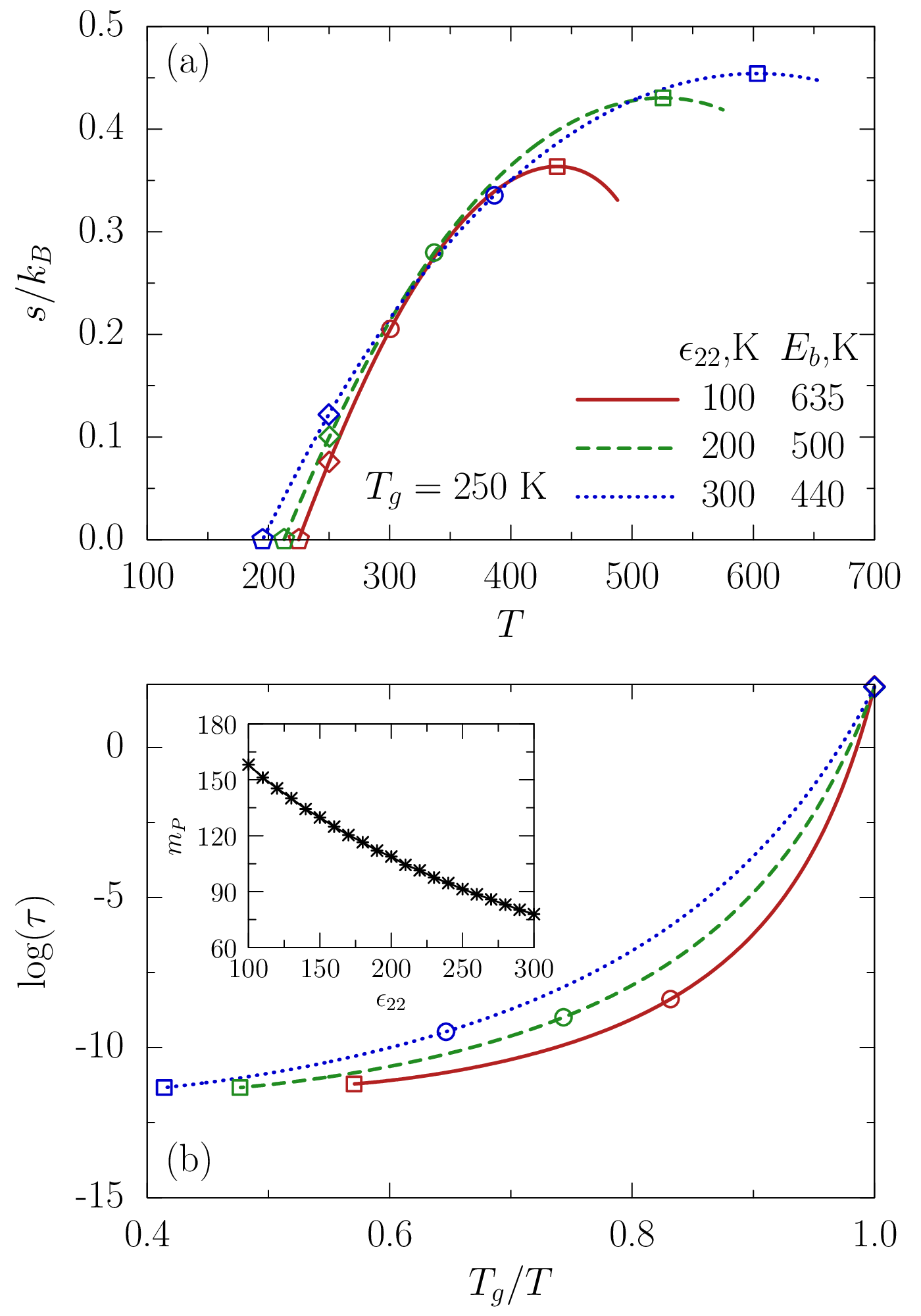}
	\caption{(a) Entropy density $s/k_B$ as a function of $T$ for several pairs of $\epsilon_{22}$ and $E_b$ that produce the same glass transition temperature of $T_g=250$ K. (b) $T_g$-scaled Arrhenius plot for the relaxation time $\tau$ for the same pairs of $\epsilon_{22}$ and $E_b$ as in part a. The inset to part b displays the variation of $m_P$ with $\epsilon_{22}$ along an iso-$T_g$ line for $T_g=250$ K. Squares, circles, diamonds and pentagons designate the positions of characteristic temperatures $T_A$, $T_I$, $T_g$ and $T_o$, respectively.}
\end{figure}

\begin{figure}[tb]
	\centering
	\includegraphics[angle=0,width=0.45\textwidth]{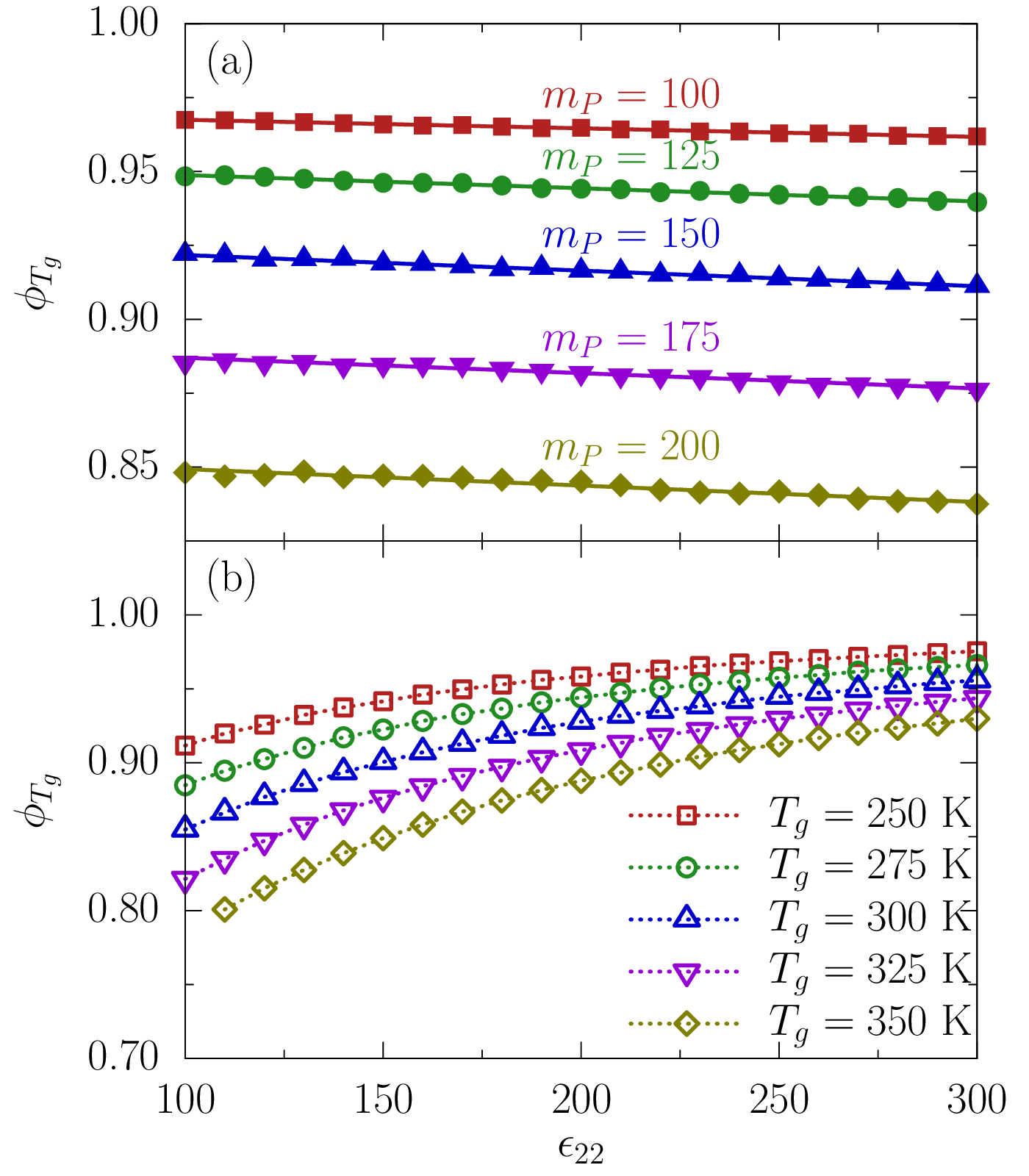}
	\caption{Polymer volume fraction at the glass transition temperature $\phi_{T_g}$ as a function of $\epsilon_{22}$ along selected iso-$m_P$ and iso-$T_g$ lines in the $\epsilon_{22}$-$E_b$ plane.}
\end{figure}

Figure 7a displays the $T$-dependence of the entropy density $s/k_B$ for different pairs of cohesive energies $\epsilon_{22}$ and bending energies $E_b$ that lie along a typical iso-fragility line for $m_P=100$. The magnitudes of the entropy density at each characteristic temperature remain approximately constant along the iso-fragility line in the $\epsilon_{22}$-$E_b$ plane. This result mirrors that observed for the monomer averaged interaction model along iso-fragility lines when using the monomer averaged interaction $\epsilon$ and bending energies $E_b$ as control parameters~\cite{Mac_47_6990}, thereby providing another evidence for the general utility of the monomer averaged interaction model.

The above result suggests that the entropy density along the iso-fragility lines may be a unique function of $T_{\alpha}/T$, i.e., the inverse temperature $1/T$ scaled by one of the four characteristic temperatures $T_{\alpha}$. This is confirmed by Figure 7b, where the entropy density along the same iso-fragility line is presented in an Angell plot. A master function approximately describes the dependence on $T_g/T$ of the entropy density along the iso-fragility lines, although the three curves in the main plot of Figure 7b slightly deviate from each other at high temperatures. The upper inset to Figure 7b also indicates that the relaxation times collapse onto a single curve when they are plotted as a function of $T_g/T$, a result that is just a consequence of the iso-fragility lines. By contrast, Figure 8 reveals that the temperature dependence of the entropy density behaves in a rather complicated fashion along the iso-$T_g$ lines. In fact, no characteristic properties are detected along the iso-$T_g$ lines. The inset to Figure 8b displays the monotonic variation of $m_P$ with $\epsilon_{22}$ along the iso-$T_g$ lines. This behavior accords with many experimental data, implying that polymers with similar $T_g$ may have very different fragilities.For example, PVC and poly($3$-chlorostyrene) (P3ClS) display similar glass transition temperatures ($T_g=352$ K for PVC and $T_g=362$ K for P3ClS)~\cite{Mac_45_8430}, but their fragilities are quite different ($m_P=191$ for PVC vs. $m_P=85$ for P3ClS)~\cite{Mac_45_8430}.

The lower inset to Figure 7b exhibits the glass transition temperature $T_g$ as increasing with $\epsilon_{22}$ along the iso-fragility lines, a conclusion that is already evident from Figure 7a. Likewise, experimental data~\cite{Mac_45_8430} indicate that the two polymers, PVA and PVC, exhibit very similar fragilities ($m_P=190$), while $T_g$ differs ($T_g=304$ K and $352$ K for PVA and PVC, respectively), illustrating that there are polymers with similar $T_g$ but different fragilities. The increase of $T_{g}$ with $\epsilon_{22}$ arises because $E_b$ increases with $\epsilon_{22}$ along the iso-fragility lines in the $\epsilon_{22}$-$E_b$ plane (Figure 6a). A simultaneous increase of $\epsilon_{22}$ and $E_b$ leads to the elevation of all characteristic temperatures. Moreover, the lower inset to Figure 7b further reveals that $T_g$ increases linearly with $\epsilon_{22}$ along the iso-fragility lines in the $\epsilon_{22}$-$E_b$ plane. Similar behavior also holds for the other characteristic temperatures (data not shown).

Our previous work~\cite{Mac_47_6990} using the monomer averaged interaction model also implies that the polymer volume fraction $\phi$ along the iso-fragility lines in the $\epsilon$-$E_b$ plane becomes a unique function of $T_g/T$ and that the volume fraction at each characteristic temperature is independent of $\epsilon$ and $E_b$. These properties fail to apply for the model of polymer melts with specific interactions, as evidenced in Figure 9a, where the polymer volume fraction at $T_g$ is shown to decrease slightly with $\epsilon_{22}$ along the iso-fragility lines in the $\epsilon_{22}$-$E_b$ plane. In addition, Figure 9b displays the polymer volume fraction at $T_g$ as increasing with $\epsilon_{22}$ along the iso-$T_g$ lines, a trend that is in agreement with that for the monomer averaged interaction model~\cite{Mac_47_6990} and can be explained by the negative correlation between $\epsilon_{22}$ and $E_b$ along the iso-$T_g$ lines (Figure 6b).

\section{Controlling Polymer Glass Formation by Specific Interactions}

Energetic heterogeneities within monomers are ubiquitous in real polymers. Even $\text{CH}_\text{n}$ groups with $\text{n} = 0,1,2$, or $3$ are generally assigned different Lennard-Jones interaction parameters in continuum model simulations~\cite{JCP_103_7156, JACS_103_335, JCP_93_4290}. However, the influence on glass formation of the energetic heterogeneities within monomers remains unclear. Because modifying the chemical structure of the backbone or side groups usually leads to changes not only in interaction energies but also in chain stiffness, the influence of energetic heterogeneities within monomers on polymer glass formation cannot readily be isolated experimentally. The specific interaction model provides an opportunity for addressing such issues. Hence, this section presents calculations in which the specific interaction energies are variables and the bending energy is held constant at $E_b=600$ K. We demonstrate that the energetic heterogeneities with monomers alone can provide an efficient means for tailoring the properties of glass-forming polymers.

\subsection{Combined Influence of Specific Interactions}

\begin{figure}[tb]
	\centering
	\includegraphics[angle=0,width=0.45\textwidth]{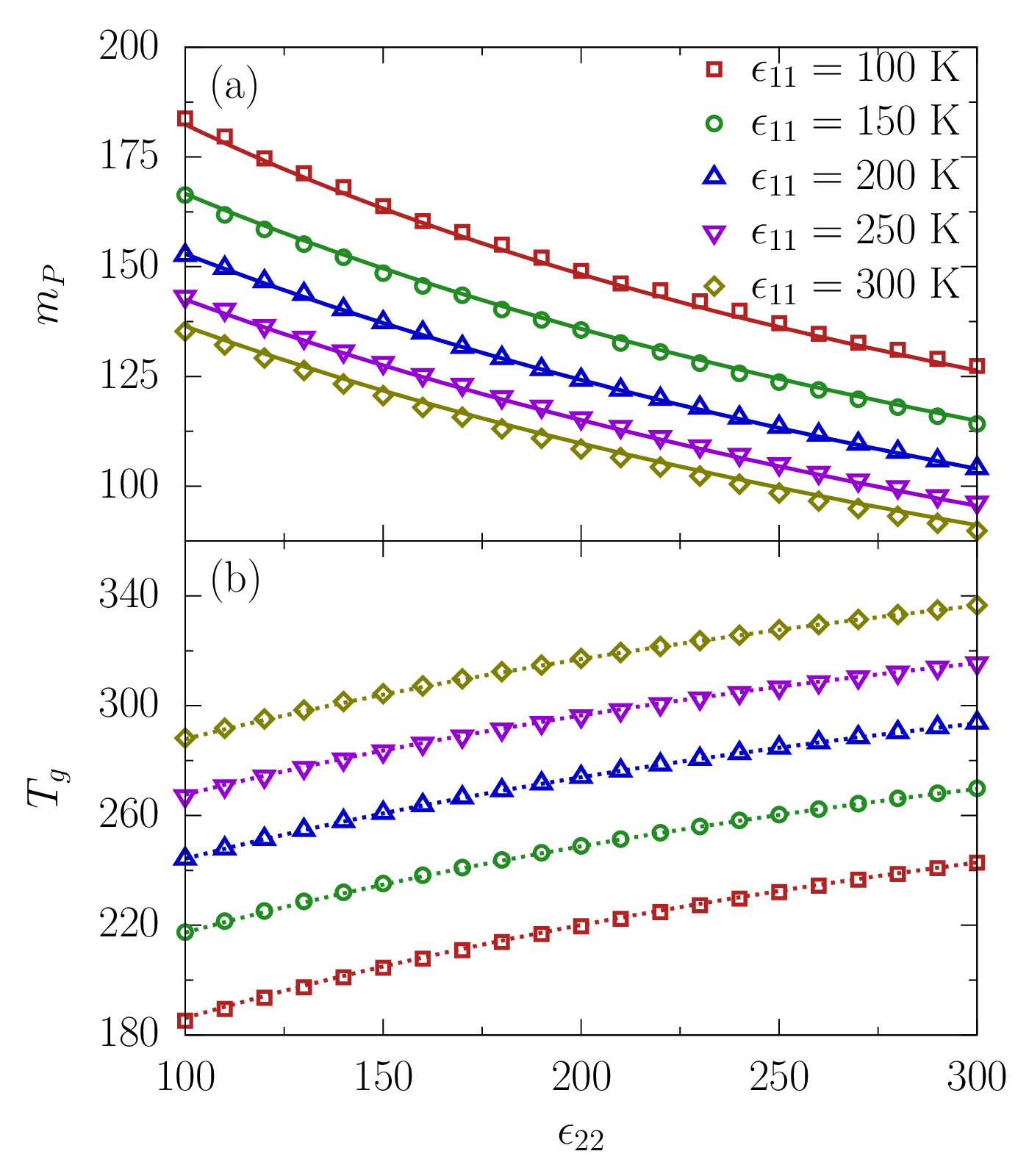}
	\caption{(a) Isobaric fragility parameter $m_P$ and (b) glass transition temperature $T_g$ as a function of the interaction energy $\epsilon_{22}$ for various interaction energies $\epsilon_{11}$. Solid lines (a) and dotted lines (b) are fits according to eqs 9 and 10 with the parameters $a_0=316.48$, $a_1=-0.872966$, $a_2=1.37215\times10^{-3}$, $b_0=0.365033$, $b_1=-3.32456\times10^{-3}$, $b_2=6.63404\times10^{-6}$, $c_0=6.43496\times10^{-3}$, $c_1=-3.25912\times10^{-5}$, $c_2=6.81742\times10^{-8}$ and $u_0=48.2178$, $u_1=0.913154$, $u_2=-9.17928\times10^{-4}$, $v_0=0.883654$, $v_1=3.21211\times10^{-3}$, $v_2=-4.65713\times10^{-6}$, $w_0=2.65577\times10^{-3}$, $w_1=7.49098\times10^{-6}$, $w_2=-1.79918\times10^{-8}$.}
\end{figure}

\begin{figure}[tb]
	\centering
	\includegraphics[angle=0,width=0.45\textwidth]{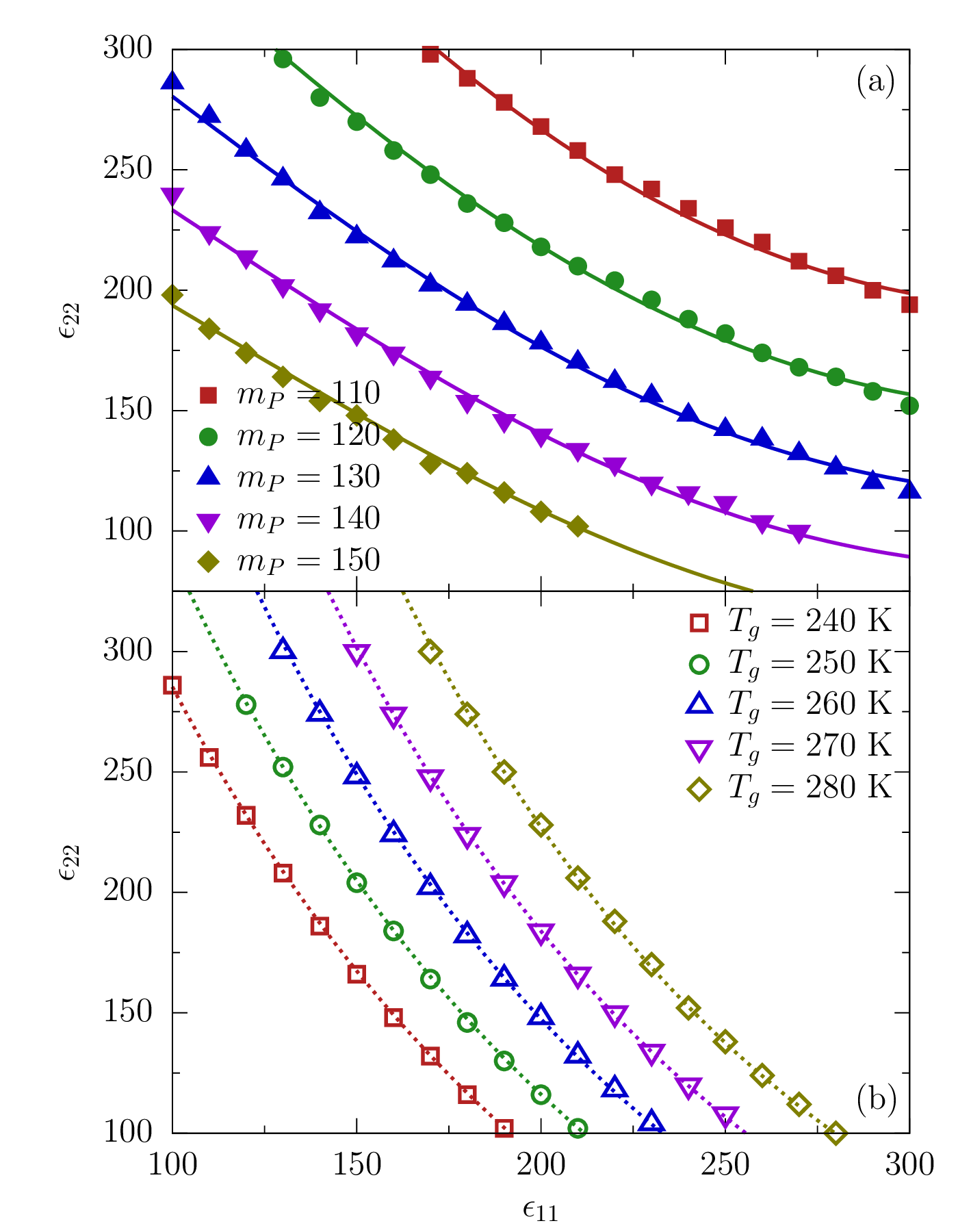}
	\caption{(a) Iso-fragility lines and (b) iso-$T_g$ lines in the plane of interaction energies $\epsilon_{11}$ and $\epsilon_{22}$ for several representative values of $m_P$ and $T_g$. Solid lines (a) and dotted lines (b) are the results of eqs 9 and 10, respectively, with the fitted parameters given in the caption of Figure 10.}
\end{figure}

Figure 10 displays the isobaric fragility parameter $m_P$ and the glass transition temperature $T_g$ as a function of side group-side group interaction energy $\epsilon_{22}$ for various backbone-backbone interaction energies $\epsilon_{11}$. The fragility parameter $m_P$ decreases with increasing either $\epsilon_{11}$ or $\epsilon_{22}$, while elevating either of the interaction energies leads to an increase in $T_g$. Apparently, the separate influence of $\epsilon_{11}$ or $\epsilon_{22}$ on $m_P$ and $T_g$ is qualitatively the same as that of the monomer averaged interaction energy $\epsilon$. Likewise, we find that two slightly different algebraic equations capture the computed combined variations of $m_P$ and $T_g$ with $\epsilon_{11}$ and $\epsilon_{22}$,
\begin{equation}
	m_P=\frac{a_0+a_1\epsilon_{11}+a_2\epsilon_{11}^2+(b_0+b_1\epsilon_{11}+b_2\epsilon_{11}^2)\epsilon_{22}}{1+(c_0+c_1\epsilon_{11}+c_2\epsilon_{11}^2)\epsilon_{22}}, 
\end{equation}
\begin{equation}
	T_g=\frac{u_0+u_1\epsilon_{11}+u_2\epsilon_{11}^2+(v_0+v_1\epsilon_{11}+v_2\epsilon_{11}^2)\epsilon_{22}}{1+(w_0+w_1\epsilon_{11}+w_2\epsilon_{11}^2)\epsilon_{22}}, 
\end{equation}
where the fitted parameters $a_{\alpha}(\alpha=0,...,2)$, $b_{\alpha}(\alpha=0,...,2)$, $c_{\alpha}(\alpha=0,...,2)$, $u_{\alpha}(\alpha=0,...,2)$, $v_{\alpha}(\alpha=0,...,2)$ and $w_{\alpha}(\alpha=0,...,2)$ are provided in the caption of Figure 10. Figure 10 demonstrates that the energetic heterogeneities with monomers alone provide additional variables for tailoring the properties of glass-forming polymers.

The iso-fragility and iso-$T_g$ lines can be similarly defined in the $\epsilon_{11}$-$\epsilon_{22}$ plane. Figure 11 displays several iso-fragility and iso-$T_g$ lines as curves of $\epsilon_{11}$ vs. $\epsilon_{22}$ for representative values of $m_P$ and $T_g$, respectively. The lines in Figure 11 display the fits obtained from eqs 9 and 10 with the parameters given in the caption of Figure 10. The features of these lines in the $\epsilon_{11}$-$\epsilon_{22}$ plane are clearly different from those in the $\epsilon_{22}$-$E_b$ plane. This result is expected since the influence of interaction energies on $m_P$ and $T_g$ differs from that of bending energy in quantitative or even qualitative ways. Figure 11 indicates that a negative correlation exists between $\epsilon_{11}$ and $\epsilon_{22}$ for both iso-fragility and iso-$T_g$ lines. Again, such a result can be explained by analyzing the combined influence of $\epsilon_{11}$ and $\epsilon_{22}$ on $m_P$ and $T_g$.

\subsection{Properties along Iso-Fragility and Iso-$T_g$ Lines}

\begin{figure}[tb]
	\centering
	\includegraphics[angle=0,width=0.45\textwidth]{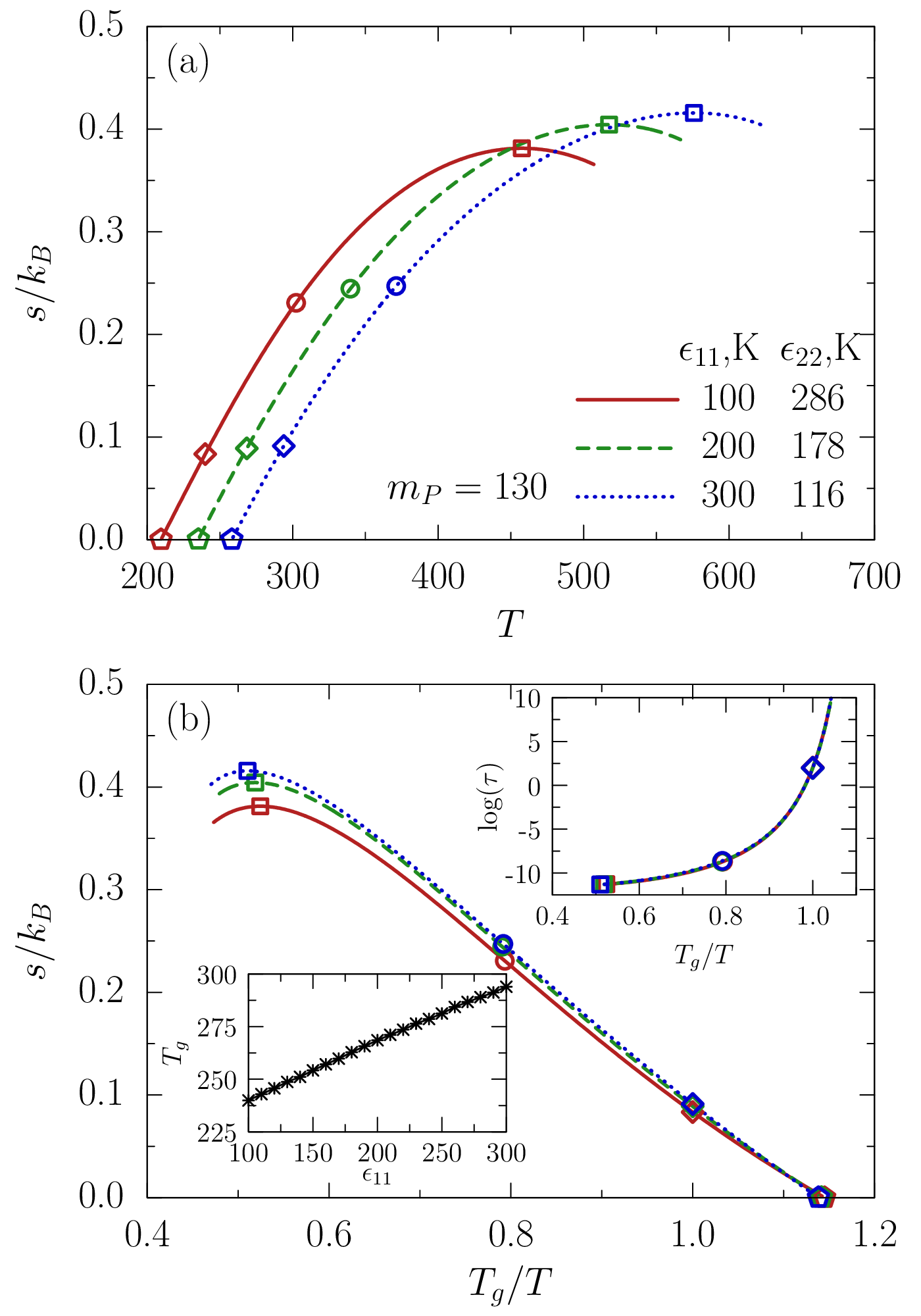}
	\caption{(a) Entropy density $s/k_B$ as a function of $T$ for several pairs of $\epsilon_{11}$ and $\epsilon_{22}$ that produce the same isobaric fragility parameter of $m_P=130$. (b) $T_g$-scaled Arrhenius plot for the entropy density $s/k_B$ for the same pairs of $\epsilon_{11}$ and $\epsilon_{22}$ as in part a. The upper inset to part b presents the $T_g$-scaled Arrhenius plot for the relaxation time $\tau$ for the same pairs of $\epsilon_{11}$ and $\epsilon_{22}$ as in part a, while the lower inset to part b depicts $\epsilon_{11}$-dependence of $T_g$ along an iso-fragility line for $m_P=130$. Squares, circles, diamonds and pentagons designate the positions of characteristic temperatures $T_A$, $T_I$, $T_g$ and $T_o$, respectively.}
\end{figure}

\begin{figure}[tb]
	\centering
	\includegraphics[angle=0,width=0.45\textwidth]{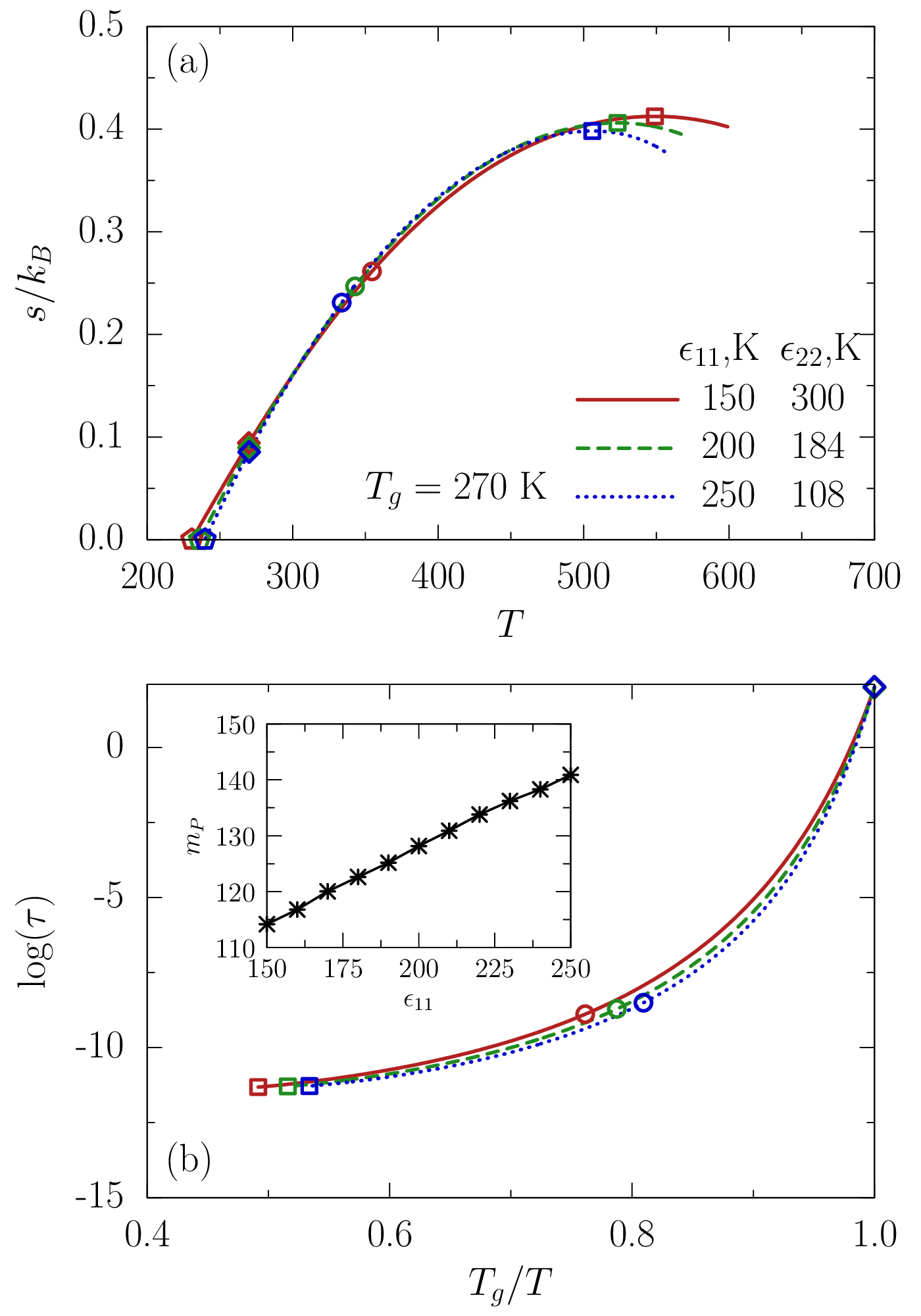}
	\caption{(a) Entropy density $s/k_B$ as a function of $T$ for several pairs of $\epsilon_{11}$ and $\epsilon_{22}$ that produce the same glass transition temperature of $T_g=270$ K. (b) $T_g$-scaled Arrhenius plot for the relaxation time $\tau$ for the same pairs of $\epsilon_{11}$ and $\epsilon_{22}$ as in part a. The inset to part b displays the variation of $m_P$ with $\epsilon_{11}$ along an iso-$T_g$ line for $T_g=270$ K. Squares, circles, diamonds and pentagons designate the positions of characteristic temperatures $T_A$, $T_I$, $T_g$ and $T_o$, respectively.}
\end{figure}

\begin{figure}[tb]
	\centering
	\includegraphics[angle=0,width=0.45\textwidth]{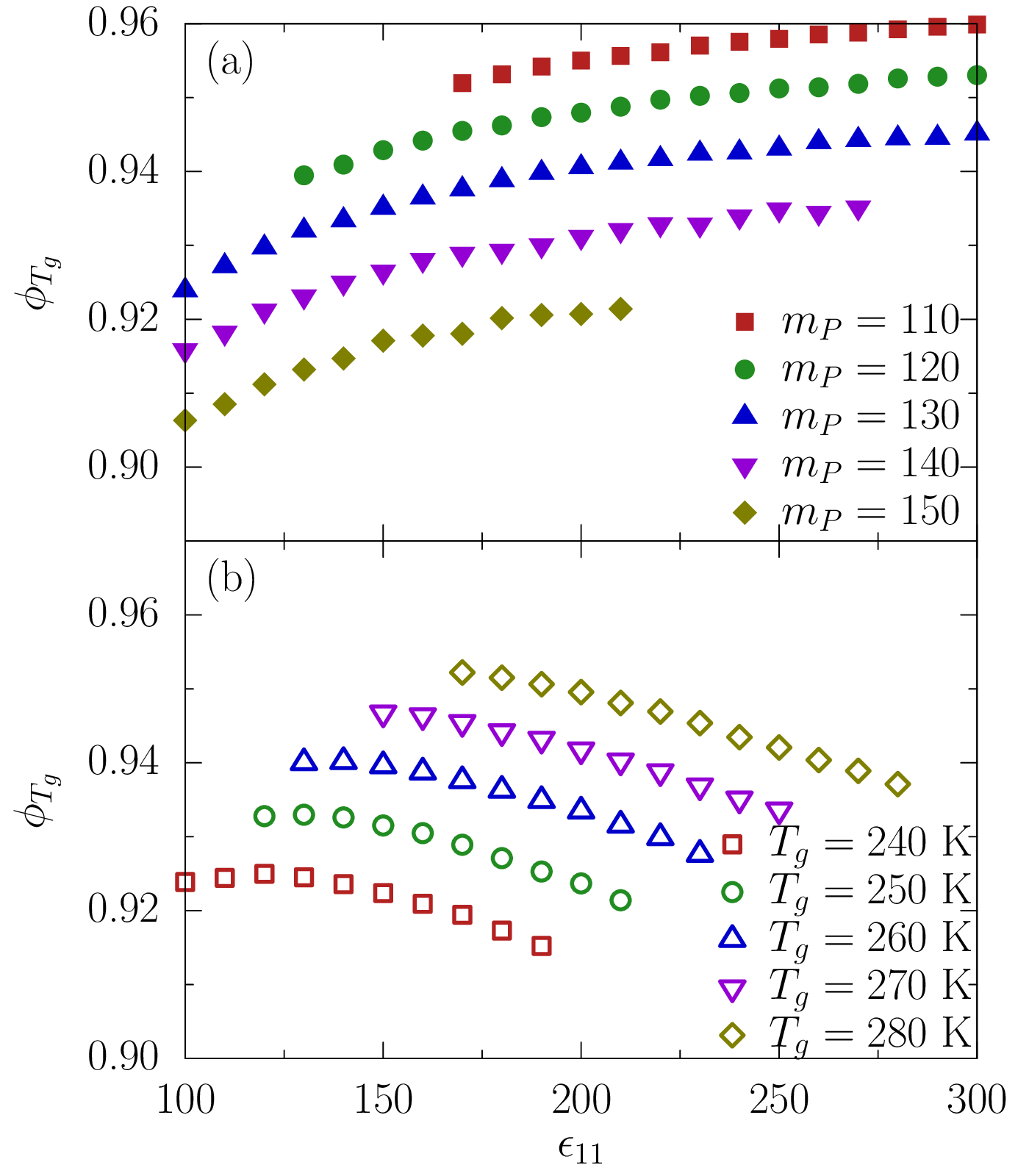}
	\caption{Polymer volume fraction at the glass transition temperature $\phi_{T_g}$ as a function of $\epsilon_{11}$ along selected iso-$m_P$ and iso-$T_g$ lines in the $\epsilon_{11}$-$\epsilon_{22}$ plane.}
\end{figure}

We now provide similar analyses for the properties along the iso-fragility and iso-$T_g$ lines in the $\epsilon_{11}$-$\epsilon_{22}$ plane and test whether the characteristic features found in the $\epsilon_{22}$-$E_b$ plane can be generalized when the bending energy is held constant.

Figure 12a displays the $T$-dependence of the entropy density for different pairs of interaction energies $\epsilon_{11}$ and $\epsilon_{22}$ that lie along an iso-fragility line for $m_P=130$. In contrast to the results in the $\epsilon_{22}$-$E_b$ plane, the magnitudes of the entropy density at characteristic temperatures ($T_A$, $T_I$, and $T_g$) clearly increase with growing $\epsilon_{11}$ or diminishing $\epsilon_{22}$ along the iso-fragility lines in the $\epsilon_{11}$-$\epsilon_{22}$ plane. This immediately suggests that the scaling of the entropy density breaks down along the iso-fragility lines in the $\epsilon_{11}$-$\epsilon_{22}$ plane (Figure 12b) and implies that the features displayed along the iso-fragility lines in the $\epsilon_{22}$-$E_b$ plane cannot be fully generalized. Nevertheless, the characteristic temperatures still increase linearly with $\epsilon_{11}$ along the iso-fragility lines in the $\epsilon_{11}$-$\epsilon_{22}$ plane (see the lower inset to Figure 12b). The upper inset to Figure 12b depicts the relaxation times as still collapsing onto a single curve when plotted as a function of $T_g/T$, and the relaxation times at the characteristic temperature $T_A$, $T_I$, or $T_g$ remain approximately constant along the iso-fragility lines in the $\epsilon_{11}$-$\epsilon_{22}$ plane. Moreover, the iso-$T_g$ lines in the $\epsilon_{11}$-$\epsilon_{22}$ plane are again not found to possess special features in the temperature dependence of both entropy density and relaxation time (Figure 13). The inset to Figure 13b exhibits the monotonic variation of $m_P$ with $\epsilon_{11}$ along the iso-$T_g$ lines in the $\epsilon_{11}$-$\epsilon_{22}$ plane, which implies that different polymers with similar $T_g$ may exhibit large variations in $m_P$ even when these polymers possess a similar degree of chain stiffness.

Figure 14a displays the polymer volume fraction at $T_g$ as clearly increasing with $\epsilon_{11}$ along the iso-fragility lines in the $\epsilon$-$E_b$ plane and, thus lacking the scaling of the volume fraction at $T_g$ obeyed by the simpler monomer averaged interaction model~\cite{Mac_47_6990}. This result emerges because van der Waals interactions exert a dominant role in determining the equation of state (EOS) of a polymer melt.~\cite{Mac_47_6990} Hence, altering both the interaction energies $\epsilon_{11}$ and $\epsilon_{22}$ is expected to induce complicated changes in the EOS, a result that also applies along the iso-$T_g$ lines in Figure 14b, which illustrates the nonmonotonic dependence of the polymer volume fraction at $T_g$ on $\epsilon_{11}$ along the iso-$T_g$ lines in the $\epsilon_{11}$-$\epsilon_{22}$ plane.

\section{Discussion}

Although energetic heterogeneities within monomers are ubiquitous in real polymers, no prior theory can address their role in polymer glass formation. The present paper extends the GET to consider the role of the energetic heterogeneities with monomers in determining the properties of glass-forming polymers. This achievement is made possible because of the recent, technically complex extension of the LCT to describe polymer melts with specific interactions~\cite{JCP_141_044909}, where three van der Waals interaction energy parameters are used to account for the energetic heterogeneities within monomers in real polymers. The greater physical realism introduced into the GET enables testing the limits of validity of the monomer averaged interaction model, which has been extensively employed to investigate the essential and physical molecular features affecting polymer glass formation~\cite{ACP_137_125, ACR_44_194, JPCB_109_21285, JPCB_109_21350, JCP_123_111102, JCP_124_064901, JCP_125_144907}, to model the glass formation of poly($\alpha$-olefins)~\cite{ JCP_131_114905}, to analyze the thermodynamic scaling of dynamics in polymer melts~\cite{JCP_138_234501}, and recently to address the two glass transitions in miscible polymer blends~\cite{JCP_140_194901, JCP_140_244905}. The computations confirm that the monomer averaged interaction model correctly captures the general trends of the variation of $m_P$ and $T_g$ with various molecular parameters. More importantly, the energetic heterogeneities with monomers alone are shown to provide an efficient way for tailoring the properties of glass-forming polymers. 

Using the extension of the CET, we further explore how the characteristic properties of glass formation in polymers can be controlled by specific interactions and chain stiffness using a minimal model containing a single bending energy. We focus on two cases. The first case employs a fixed backbone-backbone interaction energy, and the side group-side group interaction energy $\epsilon_{22}$ and bending energy $E_b$ are used as control parameters. This class of models is relevant to understanding the groups of polymers recently studied by Sokolov and co-workers~\cite{Mac_45_8430}, who explore the influence of polar interactions on polymer glass formation by comparing the results for polymers with the same backbone but different polar or hydrogen bonding side groups, the class of materials for which our model is an idealization. In the second case, the bending energy is held constant and hence, the changes in properties are solely caused by the specific interactions. Iso-fragility and iso-$T_g$ lines are defined in both cases, and we analyze the properties along the iso-fragility and iso-$T_g$ lines defined in both cases. Many features displayed by the monomer averaged interaction model also appear along those special lines in the model of polymer melts with specific interactions. However, some differences are also evident. For example, the scaling of the polymer volume fraction observed in the monomer averaged interaction model along the iso-fragility lines is absent in the model of polymer melts with specific interactions. Nevertheless, the results in the present paper clearly demonstrate the physical validity of the monomer averaged interaction model and provide evidence that tailoring molecular details such as specific interactions and chain stiffness provides an efficient route for guiding the rational design of glassy polymer materials.

\begin{acknowledgments}
We thank Jack Douglas for useful discussions and a critical reading of the manuscript. This work is supported by the U.S. Department of Energy, Office of Basic Energy Sciences, Division of Materials Sciences and Engineering under Award DE-SC0008631.
\end{acknowledgments}

\bibliography{acs}

\begin{thebibliography}{42}
\bibitem [{not()}]{note1}%
  \BibitemOpen
  \href@noop {} {}\bibinfo {note} {The range of $g$ in our calculations is much
  larger than that determined from data for polymer blends~\cite{Mac_45_1076,
  Mac_45_8861} because both $T_g$ and $m_P$ are quite insensitive to small
  changes in $g$.}\BibitemShut {Stop}%
\end{thebibliography}

\end{document}